\documentclass{article}
\title{Real continuum}
\author{O. Yaremchuk}
\date{\today}
\begin{document}
\maketitle
\begin{abstract}

Some physical consequences of the negation of the continuum
hypothesis are considered. It is shown that quantum
and classical mechanics are component parts of the multicomponent
description of the set of variable infinite cardinality.
Existence and properties of the set follow directly from the
independence of the continuum hypothesis. Particular emphasis is
laid on set-theoretic aspect.

\end{abstract}

\section{Introduction. One more aspect of the continuum problem}

Space of classical physics is the continuous set. All the physical
objects, including fields, can be removed from spatial continuum at least
theoretically. In quantum physics, the complete elimination of imbedded
structure from continuum is not possible: it is the inseparable part of the
quantum vacuum. Real spacetime continuum contains some special
microscopic ingredient. This very complicated ``insertion unit'' is, in
fact, result of size reduction of the primitive classical continuous
set over some degree of smallness. But although real space has several
absolute orders of magnitude (scales) at which different dynamical rules are
dominating, the basic structure of mathematical continuum (the set of all
real numbers) does not have any criterion of size.

It is well known that any interval of continuum has the same number of
points as the entire set of the real numbers.
Moreover, arbitrarily small, even infinitesimal, interval of the real line
has the same number of points as all the continuous universe of
any number of dimensions.

In order to make properties of formal continuum conform to variable
properties of real continuum and avoid the insertion structure, we shall
correct this implausible superhomogeneity of the set of the real number.

Equivalence of all continuous intervals (without loss of generality we
shall consider one-dimensional case) should be replaced by the
following realistic dependence pattern of cardinality of an interval on
its size: When the interval is large enough, its cardinality is close to
cardinality of continuum (this is just mere establishment of the fact),
i.e., all the intervals of space, down to some degree of smallness,
are practically equipotent. Decrease of the number of points of the
intervals is imperceptible.

This means that all such intervals have regular length, since
length implies one-to-one correspondence between the points of any
of the intervals $l_{large}$ and the set of the real numbers $R$ and, thus,
may be regarded as manifestation of the equivalence
\begin{equation}\label{length}
l_{large}\leftrightarrow R.
\end{equation}

When difference between cardinality  of an interval and cardinality of
continuum become substantial, it should adversely affect length of the
interval. Strictly speaking, length, as the establishment of equivalence
$l_{large}\leftrightarrow R$, should vanish.
Formally, a small interval such that $|l_{small}|<|R|$ turns to point.
But it is natural to expect existence of a transition region:
length of the noticeably non-continuous interval must show some irregularity,
Measurements (direct or indirect) cannot give a unique stable real number.

Since any infinite set should be equivalent to its proper subset,
infinite number of points decreases by steps: only some infinite
``portion'' of points changes the total number (makes the set
non-equivalent to the initial one). By this reason, we can get
only infinite number of points as a final result (finite set is not
equivalent to any of its proper subset). In other words, real continuum
is not infinitely divisible, i.e., there exists the infinite minimal set
(greatest lower bound) instead of the minimal length.

If the infinite number of points of a continuous interval can decrease and
the set of the points of sufficiently small interval becomes non-equivalent
to the set of the real numbers, then the continuum hypothesis is false,
i.e., the above assumption may be regarded as a form of the negation of the
continuum hypothesis (CH) which seemingly contradicts to its
independence. Note that the independence of CH can be established
only in the framework of certain formal system, whereas the continuum
problem had been stated before creation of any axiomatic set theory.
Informally, the independence is not a final solution. Such a solution
should either determine status of the set of intermediate cardinality
or show that the continuum problem is meaningless.

Fortunately, there is a unique status of the intermediate set consistent
with the independence of CH: since, by definition, continuum $R$ should
contain the subset of intermediate cardinality $M$ such that
$|N|<|M|<|R|$, where $N$ is a set of the natural numbers, the
independence of CH means that for any real number $x\in R$ the
statement $x\in M\subset R$ is undecidable, i.e., the intermediate
subset cannot be extracted from continuum. In other words, since
non-existence of the set clearly contradicts the independence of CH,
the only possible understanding is inseparability of the subset.
Reason for this confinement should be investigated.

However, postponing investigation of the reason, we can use the
independence of CH in order to get quite rigorous result.

\section{From the continuum problem to path integrals}

\subsection{Thesis}

The latent status is the only definite status of the
set of intermediate cardinality that is consistent with the generally
accepted solution of the continuum problem. However, it is necessary to
know that standard Zermelo-Fraenkel set theory (ZF) gives correct
description of the notion of set, i.e., we need set-theoretic analog of
Church's thesis in order to be sure that we have reliable solution of the
continuum problem independent of the concrete formalization of the concept
of set. 

\subsection{Maps}

Consider the maps of the intermediate set $I$ to the sets of real numbers
$R$ and natural numbers $N$:
\begin{equation}
N\gets I\to R.
\end{equation}

Let the map $I\to N$ decompose $I$ into
the countable set of mutually disjoint
infinite subsets: $\cup I_n=I$ ($n\in N$).
Let $I_n$ be called a unit set. All members of $I_n$
have the same countable coordinate $n$.

Consider the map $I\to R$.
By definition, continuum $R$ contains a subset $M$ equivalent to $I$,
i.e., there exists a bijection
\begin{equation}
f:I\to M\subset R.
\end{equation}
This bijection reduces to separation of the intermediate
subset $M$ from continuum. For example, the separation of
three real numbers is equivalent to the bijection
${(1,2,3)\to R}$. If we do not use any rule for the separation,
we get the random (arbitrarily chosen) numbers
${(r_1,r_2,r_3)}$.
This randomness is not of principle because there are
many rules for separation of three numbers as well as
of any finite or countably infinite number of the real numbers.
But in the case of the intermediate set, we, in principle,
do not have a rule for separation of any subset with this
number of members, since any separation rule for such a subset
that can be expressed in ZF is a proof of existence of the
intermediate set and, therefore, contradicts the independence
of the continuum hypothesis. 

Thus we, in principle, do not have a rule for
assigning a definite real number to an arbitrary
point $s$ of the intermediate set. Hence, any
bijection can take the point only to a random real number.

\subsection{Intermediate set}

This does not mean that the intermediate set consist of random
numbers or that the members of the set are in any other sense
random. Each member of the set of intermediate cardinality equally
corresponds to all real numbers until the mapping has performed
operationally. After the mapping, the concrete point gets the random
real number as its coordinate in continuum, i.e., we get the
probability $P(r)dr$ of finding the point $s\in I$ about $r$.

The independence of the continuum hypothesis is proved by construction
of models of ZF with and without the intermediate set. In contrast to
the model without the set which is the ``smallest set theory'' consisting
only of constructible sets,  the model with the ``set of intermediate
cardinality'' is some unnatural extension of set theory or rather its
distortion.  Of course, the real intermediate set is not constructed.
These models establish that the existence of the set of intermediate
cardinality does not affect the formalized properties and interrelations
of sets. Thus one cannot state that the intermediate set does not exist but
the effect of its presence is absent. The set is ``ZF-imperceptible''.

\subsection{Coordinates}

Each member of the set of intermediate cardinality equally
corresponds to all real numbers until the mapping has performed
operationally. After the mapping, a concrete point gets random
real number as its coordinate in continuum.
Thus the point of the intermediate set has two coordinates:
a definite natural number and a random real number:
\begin{equation}\label{s}
s:(n,r_{random}).
\end{equation}
Only the natural number coordinate gives reliable
information about relative positions of the points
of the set and, consequently, about size of an interval.
But the points of a unit set are indistinguishable.

\subsection{Transition}

Consider probability $P(b,a)$ of finding the point $s$ at $b$
after it was found at $a$.
The interval $(a,b)$ defines parameterization
of the coordinates of the point $s$. Let the parameter
be denoted by $t$:
\begin{equation}\label{s(t)}
s(t):[n(t),r(t)],
\end{equation}
where $a<t<b$, $a=t_a=r(t_a),\,b=t_b=r(t_b)$.
Note that $n=n(r)$ does not exist: since $r=r(n)$ is random number,
the inverse function is meaningless. 

We have the right to consider the behavior of the point between
$a$ and $b$ and to identify this parameter with time.

Since the point at any $t$ corresponds to all real numbers
simultaneously, it corresponds to all continuous random
sequences of the real numbers (paths) $r(t)$. The elemental
events are not mutually exclusive and, therefore, 
\begin{equation}
P(b,a)\ne\!\!\sum_{all\,r(t)} P[r(t)],
\end{equation}
where $P[r(t)]$ is the probability of finding the point $s$
at any $t$ on some arbitrary path (a continuous sequence
of random real numbers) $r(t)$.

\subsection{Probability}

We cannot compute the probability $P(b,a)$ in the
ordinary way, i.e., by summing or integration of $P[r(t)]$.
In order to overcome this obstacle, it is most natural
to introduce some additive functional $\phi[r(t)]$
such that
\begin{equation}
P[r(t)]={\cal P}\{\phi[r(t)]\}
\end{equation}
and
\begin{equation}
P[b,a]={\cal P}(\!\!\sum_{all\,r(t)}\!\!\phi[r(t)]).
\end{equation}
In other words, we shall compute the non-additive
probability from the additive functional by a simple
rule. It is clear that the dependence should be
non-linear:
\begin{equation}
{\cal P}(\!\!\sum_{all\,r(t)}\!\!\phi[r(t)])\ne\!\!
\sum_{all\,r(t)} \!\!{\cal P}\{\phi[r(t)]\}].
\end{equation}

We may choose the dependence arbitrarily. The simplest
non-linear dependence is the square dependence:
\begin{equation}
{\cal P}[r(t)]=|\phi[r(t)]|^2.
\end{equation}

The function $\phi$, obviously, depends on $n(t)$.
Since at any $t$ the point equally corresponds to all the
real numbers, it equally corresponds to all $r(t)$. This
symmetry is of principle because it follows directly from
the independence of the continuum hypothesis.
As a result of the symmetry, all $r(t)$ are equiprobable:
$P[r(t)]$ does not depend on $r(t)$, therefore,
modulus of $\phi$ is constant and $n(t)$ may appear only
in its phase. At the same time, it is necessary to ensure
invariance of the $P[r(t)]$ under shift in $N$:
\begin{equation}
|\phi[r(t),n(t)]|^2=|\phi[r(t),n(t)+const]|^2.
\end{equation}
Hence, the function $\phi$ is of the following form:
\begin{equation}
\phi[r(t)]=const\,e^{2\pi i\,F[n(t)]},
\end{equation}
where $F[n(t)]$ is some real-valued additive functional of $n(t)$.

It is clear that $F[n(t)]$ is directly proportional to the other
additive functional of $n(t)$: the length $m$ of the countable path
$n(t)$.
Let us put:
\begin{equation}
F[n(t)]=m.
\end{equation}
Then we get
\begin{equation}
\phi[r(t)]=const\,e^{2\pi im}
\end{equation}
and
\begin{equation}\label{pathsum}
P(b,a) = |\!\!\sum_{all\,r(t)}\!\!\mbox{const}\,e^{2\pi im}|^2,
\end{equation}
i.e., the probability $P(a,b)$ of finding the point $s$ at $b$
after finding it at $a$ satisfies the conditions of Feynman's approach
(section 2-2 of \cite{Feynman}) for $S/\hbar=2\pi m$. 
Therefore,
\begin{equation}
P(b,a)=|K(b,a)|^2,
\end{equation}
where $K(a,b)$ is the path integral (2-25) of \cite{Feynman}:
\begin{equation}\label{pathint}
K(b,a)=\int_{a}^{b}\!e^{2\pi im}{\cal D}r(t).
\end{equation}
Since Feynman does not essentially use in Chap.2 that $S/\hbar$ is just
action, the identification of $2\pi m$ and $S/\hbar$ may be postponed.

\subsection{Principle of least action, quantum of action, and mass}

In section 2-3 of \cite{Feynman} Feynman explains how
the principle of least action follows from the dependence
\begin{equation}\label{sum}
P(b,a)= |\!\sum_{all\,r(t)}\!\!\mbox{const}\,e^{(i/\hbar)S[r(t)]}|^2.
\end{equation}
This explanation may be called ``Feynman's correspondence principle''.
We can apply the same reasoning to Eq.(\ref{pathsum}) and,
for very large $m$, get ``the principle of least $m$''.
This also means that for large $m$ the point $s$ has a definite
stationary path and, consequently, a definite continuous coordinate.
In other words, the interval of the intermediate set with the large countable
length $m$ is sufficiently close to continuum (let the interval be
called macroscopic), i.e., cardinality of the intermediate set depends
on its size.

For sufficiently large $m$,
\begin{equation}
F[n(t)]=\int_a^b\!\! dm(t)=\int_a^b\!\! \frac{dm(t)}{dt}\,dt.
\end{equation}
Note that $m(t)$ is a step function and its time derivative is almost
everywhere exact zero. But for sufficiently large increment
$dm(t)$ the time derivative  ${\frac{dm}{dt}=\dot{m}(t)}$
makes sense as non-zero value.

The function $m(t)$ may be regarded as some function of
$r(t)$: $m(t)=\eta[r(t)]$. It is important that $r(t)$ is not
random in the case of large $m$. Therefore,
\begin{equation}\label{f}
\int_a^b\!\!dm(t)=\int_a^b\!\frac{d\eta}{dr}\,\dot{r}\,dt=\min,
\end{equation}
where $\frac{d\eta}{dr}\,\dot{r}$ is some function of $r$, $\dot{r}$,
and $t$. This is a formulation of the principle of least action
(note absence of higher time derivatives than $\dot{r}$), i.e.,
large $m$ can be identified with action.

Recall that this identification is valid only for very large $dm=\dot{m}dt$,
i.e., for sufficiently fast points. In fact, this is a qualitative leap:
action is not the length of the countable path but some new function.
We get a new characteristics of the point and a new law of its motion.

Since the value of action depends on units of measurement, we need
a parameter $h$ depending on units only such that
\begin{equation}
hm=\int_a^b\!\! L(r,\dot{r},t)\,dt=S.
\end{equation}

Finally, we may substitute $S/\hbar$ for $2\pi m$ in Eq.(\ref{pathint})
and regard our consideration as a natural extension of Feynman's formulation
of quantum mechanics.

The original Feynman's approach becomes more consistent with this extension
because there is no need in physically meaningless segments of straight
line or sections of the classical orbit between the points of the partition
Eq.(2-19), Fig. 2-3 (\cite{Feynman}) by which Feynman constructs the sum
over paths. There is also no need in existence of action from the very
beginning.

Note that if time rate of change of cardinality (i.e., of the countable
coordinate) is not sufficiently high, action vanishes: $\dot{m}(t)$ and,
consequently, ${dm=\dot{m}(t)dt}$ is exact zero. This may be understood
as vanishing of the mass of the point. Formally, mass is a consequence of
the principle of least action: it appears in the Lagrangian of a free
particle as its peculiar property \cite{mech}. Thus mass is somewhat
analogous to air drag which is substantial only for sufficiently fast
bodies.

\section{Extra descriptions and extra dimensions}

\subsection{Intervals}

If we reduce some interval of real continuum to the order of
smallness, at which decrease in its cardinality is appreciable, we
automatically reveal discrete properties of the interval:
due to equivalence of any infinite set to its proper
subset, cardinality decreases in steps, i.e., the interval
becomes less similar to continuum (instability of its length)
and more similar to the countable set (it gets one more length
expressed by natural number).

Thus we get three kinds of the intervals:

large continuous intervals that have length as manifestation of their
equivalence to the set of the real numbers;

insufficiently large submicroscopic intervals whose lengths are therefore
unstable (the interdependence between instability of the interval length
and its natural number length is the content of quantum mechanics);

small non-continuous microscopic intervals without length which are,
actually, composite points.

\subsection{Dimensionality}

In order to keep inside certain cardinality, a shift should also has
this cardinality. Therefore, one-dimensional intermediate axis splits
into, at least, three non-equivalent ``subaxes,'' i.e., immiscible
substructures. The complete description is three-dimensional.

In this case, dimensionality is a classification of cardinalities.
The classification with respect to length is the roughest (macroscopic)
estimate of cardinality (yes, no, unstable). More precisely, length is an
indication of degree of saturation of cardinality: saturated (continuum),
unsaturated, close to saturation, respectively.

Saturation of cardinality is important because of the following paradox
which is important for understanding of mechanical motion:
when a point moves with high countable speed that may be regarded as a
continuous variable, saturated cardinality of the path and its time rate
of change are really constant. Cardinality does not change as in the case
of ``countably motionless'' (massless) point. Very fast ``vertical'' motion
turns to ``horizontal''.

Classical mechanics gives only one spatial dimension: the continuous
coordinate. The independent natural number coordinate is replaced by the
functional of the continuous coordinate and its time derivative
(degeneration).

Quantum mechanics gives two coordinates: natural number and random real
number.

The proper microscopic description does not give extra dimensions if the
proper microscopic intervals are regarded as points. These composite points
take part in classical and quantum-mechanical descriptions. However, since
the microscopic intervals are essentially non-equivalent, they themselves
are immiscible and form a quantity of different objects described by a
hierarchy of theories. Therefore, description of the structure and
transmutation of the intervals needs additional dimensions down to the
single unit set. But these dimensions should manifest themselves rather as
qualitative properties (charges) of the points (in other words, they are
inherently ``compact'').

Thus we get three spatial dimensions (one macroscopic and two microscopic)
and time in the one-dimensional case. It is interesting to note that in the
three-dimensional case it gives ten spacetime dimensions just like in string
theory. We also expect some unknown number of extra dimensions ``inside the
point''.

Since the description of the one-dimensional intermediate set consist
of ``sections'', which are on equal footing, the particular main laws,
directions, and dimensions of the ``sections'' are equally valid. Thus
we get parallel descriptions. These descriptions relate to different
immiscible substructures of real continuum.

It is quite clear, regretfully post factum, that classical and quantum
mechanics are mutually irreducible by any correspondence principle.
Reducing Plank constant to zero, one cannot get classical mechanics.
One can only make some operators commute.

Feynman's correspondence principle does not reduce classical mechanics
to quantum mechanics but separates their fields of application.
These fields are not only different scales of the same space
(there are well-known macroscopic quantum phenomena).
The distinction goes further because, due to non-equivalence of the
subsets of different cardinalities, these subsets are closed under
different equivalence relations (symmetry transformations) that leads to
effect of separate dimensions and, consequently, directions. Therefore,
these non-equivalent structures (ruled by different laws) become
autonomous and immiscible.

\section{Set theory  and real continuum}

\subsection{Fission vs. construction}

It is worthwhile to pay attention to the way of obtaining sets by
fission of continuum. According to P. Cohen, continuum ``can never
be approached by any piecemeal process of construction'' \cite{Cohen},
therefore, it may be stated that members and subsets of continuum
obtained by such a process are not true but at best imitation, i.e.,
the true members and subsets should be extracted from continuum itself
by its fission.

\subsection{Cardinality as a property}

According to the separation axiom scheme, for any set and for any
property expressed by some formula there exists a subset of the set,
which contains only members of the set having the property.
From the independence of CH it follows that we cannot express
any property of the members of the intermediate set, i.e., any property
we can formulate implies separation of either countable or continuous
subset of continuum. This fact is the reason of the inseparability of the
set of intermediate cardinality. The simplest way out is unexpected:
to regard infinite cardinality itself as a property of the set members.
Breaking real continuum, in which cardinality of any part depends on the
size of this part, into intervals of lower cardinalities, we should
consider each intermediate cardinality as an elementary property (``charge,''
power) of the corresponding interval.

Then we get one more aspect of the continuum problem: how many different
properties (non-equivalent infinite fragments) can be derived from continuum?

From this standpoint, members of continuum are not real numbers,
subsets of $N$, or zero length points but continuous intervals.
The representation of the real numbers as nonterminating decimals
implies infinite process of fission. On the contrary, intervals
of continuum can be obtained by primitive finite procedures.

\subsection{Intuition of continuum}

We, obviously, have intuition of continuum but this intuition is not used
in set theory perhaps because it does not coincide with the set of the real
numbers: equivalence of an arbitrarily small interval to the entire set of
the real numbers is clearly counter-intuitive.

At first sight, ``axiom of continuum,'' stating existence of the
structureless continuous whole, and formal scheme for its fission into
equivalent and non-equivalent parts seem unavoidable in order to
complete (balance) set theory: Zermelo-Fraenkel set theory has tools
only for construction of sets. However, it is more important to have
consistent factual picture independent of formalization.
Formal results often need interpretations which sometimes constitute
more difficult problems than formal solutions themselves. For instance,
the continuum problem is much clearer when it is stated informally.
It is interesting that formalists still are not sure that the problem and
the problem of size of continuum generally make sense.

For the twentieth century, which was the century of search
for the formal unity (universal formalization), a great number
of statements in different areas of mathematics had been
shown to be independent. This is quite explainable, taking
into account that the present mathematics is inevitably
macroscopic. Statements and concepts touching upon essentially
microscopic aspects should be either independent, which
shows absence of necessary information, or contradictory,
which indicates attempt to unify distinct fields of
reality. Sets are real objects and hardly can serve as
unchangeable ``mental units''.
Set theory with the intermediate set cannot be separated from,
at least, microscopic (quantum) reality because set-theoretic
and even logical notions appear to be connected with fine
spatial structure.

Fundamental physics may be called ``study of real continuum''.
The G\"odel's incompleteness theorems, applied to the study, can be
interpreted as impossibility of a unique unified theory of everything.
The second incompleteness theorem points to the hierarchical structure of
fundamental theories. A correct theory is not a limiting case of the next
more exact theory. The correct theories form some structure which is
directly related to the structure of continuum itself.

Since the complete description consist of interpenetrating parts
governing by different rules, one can get formal contradiction as a
real conflict of correct descriptions. In this case, elimination of
contradictions in order to get consistent unified formal picture
is inadmissible.

\subsection{Cardinality and structure}

Cantor's opinion that cardinality of a set is independent of nature and
properties of its members is still regarded as indisputable.
However, this is neither axiom nor theorem but only an observation on finite
sets by default imposed on infinite ones. Note that, from some number of
line segments, one can form the most complex structure (e.g. polygon) which
may serve as a unique characteristic of the number, i.e., besides one-to-one
correspondence, finite cardinality can be characterized by some structure.

Unlike a finite set, an infinite collection of members cannot be in
disordered state or arbitrarily arranged. Since an infinite set is
equivalent to its proper subset, it may be stated that any infinite
set necessarily forms some symmetrical structure (asymmetrical
arrangement is impossible).

Note that, the most symmetrical arrangement is the most probable one because
such an arrangement has the greatest number of equivalent (symmetrical)
states (the number of ways in which the arrangement can be produced:
``thermodynamical probability'' of the arrangement of the infinite set).
Physically, this means that only the most symmetrical arrangement is stable,
i.e., any infinite set forms the most symmetrical structure by itself.
It is clear that symmetries (equivalence relations) of non-equivalent
sets should be different. Hence, an infinite set can be characterized
by type of symmetry and ``charge'' of its members. 
 
We can formulate the following rule: any infinite set tends to form the
most symmetrical structure determined by its cardinality.

In the world of finite sets, we need more bricks for a big palace than for
a small house. The most complex building can play the role of the primitive
symbol of the corresponding number of bricks. In the world of infinite sets,
the role of such a symbol plays, figuratively, the smallest cabin, i.e., the
simplest (most symmetrical) structure.

Whole continuum is regarded as elementary structureless object, to a certain
degree, complementary to the empty set in present set theory which is
obviously also without structure.

Absence of structure explains absolute homogeneity of the formal
construction (the set of all real numbers) which is unconditionally
identified with continuum.

In the case of macroscopic system, the rule of maximum symmetry
works like the law of entropy increase: since any macroscopic system
occupies continuous region, it tends to reproduce the absolutely
homogeneous structureless continuous whole by disintegration of all
macroscopic structures as inhomogeneities and making chaos.
However, submicroscopic and proper microscopic objects successfully avoid
this law because decrease of cardinality automatically entails structures,
i.e., final states of these objects are structured (but non-stationary
because symmetrization in multisructured system leads to dynamics).
Hence, contrary to our expectation, we get increase of complexity
of smaller objects. In other words, primitive spatial continuum really
contains more complicated ``insertion units;'' the whole is simpler than
its component parts. This conclusion is very strange indeed. Figuratively,
complex microscopic structures are cut out of the whole ``piece of wood''
and then assembled into constructions (``Pinoccio making method'').

Thus even global tendencies (``fates'') of macroscopic, submicroscopic,
and microscopic subworlds are different. For instance, atoms and particles
never get old, while large molecules, e.g. protein, are already subjected
to aging. Recall also that, in pure quantum systems, chaos, in classical
sense, is absent. Quantum chaos can be defined only for semiclassical
systems and this is rather theoretical possibility than phenomenon needing
obligatory explanation.

Consequently, all the structures in our continuous universe have microscopic
origin and are supported by the microscopic processes. Note that the only
pure macroscopic object is classical vacuum which is really structureles.
Geometry (and fields) requires presence of microscopic structures (matter).
On the contrary, microscopic point-like objects are complex and can contain
much more information than it is supposed.

If intermediate cardinalities are regular, then fission is irreversible:
one cannot restore continuum or any intermediate interval as the union
of the intervals of lower cardinalities. It is plausible because such a
restoration is not a mere union of sets but synthesis of more homogeneous
structure of higher cardinality from structures of lower cardinalities
(``regeneration'').

It is interesting that, practically, abstract set-theoretic regularity gives
the notion of space itself in its visual sense (emptiness): structures, i.e.,
objects of smaller cardinalities (matter), cannot fill all the continuous
superset. 

\subsection{Open-closed duality}

Until the intermediate set is large enough to be regarded as
continuous, the most symmetrical structures it can form are
closed structures (loops): A small intermediate interval consist
of a small finite number of unit sets. A finite set is not
equivalent to any of its proper subsets, i.e., it has no natural
symmetries (self-coincident moves) but the interval, as an infinite set,
should take the most symmetrical form. The only way to get a natural
symmetry is to form a loop.
Note that the least number of unit sets for a loop is three.

It is also natural to expect existence of transition region,
where closed interval is not stable enough (open-closed duality).
Only sufficiently continuous interval can form stable open structure.
However, such intervals are miscible with the continuum
(the stable continuous intervals are members of continuum).
Thus most of the strings should be closed; only more or less unstable
open strings can be observed.

\subsection{Real and virtual subsets}

Since formal continuum (the set of the real numbers) differs from real
one, the reason of the inseparability of the intermediate subset in the set
of the real numbers is different from that in real continuum.
Whereas the large intermediate set really contains such subset,
the natural structure of formal continuum is only a carrier medium for the
unstable manifestations of the members of the intermediate set.
In the set of the real numbers, unstable members are artificial objects:
they are not component parts but only possible formations in the structure
of the set. However, one cannot state that exact continuum does not
contain the intermediate subset, albeit its presence is rather virtual.

\end{document}